\documentclass[sigconf]{acmart}

\usepackage{booktabs} % For formal tables

\usepackage{lipsum} 
\usepackage{balance}
\usepackage{paralist}
\usepackage{filecontents}
\usepackage{etoolbox}
\usepackage{mathrsfs}
\usepackage{bm}
\usepackage{setspace}
\usepackage{amssymb}
\usepackage{graphicx}
\usepackage{caption}
\usepackage{subcaption}
\usepackage[ruled]{algorithm2e}
\usepackage{multirow}

\copyrightyear{2018} 
\acmYear{2018} 
\setcopyright{acmcopyright}
\acmConference[SIGIR '18]{The 41st International ACM SIGIR Conference on Research and Development in Information Retrieval}{July 8--12, 2018}{Ann Arbor, MI, USA}
\acmBooktitle{SIGIR '18: The 41st International ACM SIGIR Conference on Research and Development in Information Retrieval, July 8--12, 2018, Ann Arbor, MI, USA}
\acmPrice{15.00}
\acmDOI{10.1145/3209978.3209986}
\acmISBN{978-1-4503-5657-2/18/07}

\begin{document}
\title{Unbiased Learning to Rank with Unbiased Propensity Estimation}

\iffalse
\author{~}
%\authornote{Dr.~Trovato insisted his name be first.}
\orcid{~}
\affiliation{%
	\institution{~}
	\streetaddress{~}
	\city{~} 
	\state{~} 
	\postcode{~}
}
\affiliation{%
	\institution{~}
}
\email{ }
\fi

\author{Qingyao Ai}
%\authornote{The secretary disavows any knowledge of this author's actions.}
\affiliation{%
	\institution{CICS, UMass Amherst}
	%\streetaddress{P.O. Box 1212}
	\city{Amherst} 
	\state{MA} 
	\country{USA}
	%\postcode{01003-9264}
}
\email{aiqy@cs.umass.edu}
\author{Keping Bi}
%\authornote{The secretary disavows any knowledge of this author's actions.}
\affiliation{%
	\institution{CICS, UMass Amherst}
	%\streetaddress{P.O. Box 1212}
	\city{Amherst} 
	\state{MA} 
	\country{USA}
	%\postcode{01003-9264}
}
\email{kbi@cs.umass.edu}
\author{Cheng Luo}
\affiliation{%
	\institution{DCST, Tsinghua University}
	%\streetaddress{P.O. Box 1212}
	\city{Beijing} 
	\country{China} 
	%\postcode{100190}
}
\email{luochengleo@gmail.com}
\author{Jiafeng Guo}
\affiliation{%
	\institution{ICT, Chinese Academy of Sciences}
	%\streetaddress{P.O. Box 1212}
	\city{Beijing} 
	\country{China} 
	%\postcode{100190}
}
\email{guojiafeng@ict.ac.cn}
\author{W. Bruce Croft}
%\authornote{The secretary disavows any knowledge of this author's actions.}
\affiliation{%
	\institution{CICS, UMass Amherst}
	%\streetaddress{P.O. Box 1212}
	\city{Amherst} 
	\state{MA} 
	\country{USA}
	%\postcode{01003-9264}
}
\email{croft@cs.umass.edu}

\iffalse
\author{Qingyao Ai$^1$,Keping Bi$^1$, Cheng Luo$^2$, Jiafeng Guo$^3$,  W. Bruce Croft$^1$}
%\authornote{The secretary disavows any knowledge of this author's actions.}
\affiliation{%
	\institution{$^1$College of Information and Computer Sciences, University of Massachusetts Amherst}
	%\streetaddress{P.O. Box 1212}
	\city{Amherst} 
	\state{MA} 
	\country{USA}
	%\postcode{01003-9264}
}
\email{{aiqy, kbi, croft}@cs.umass.edu}
\affiliation{%
	\institution{$^2$DCST, Tsinghua University}
	%\streetaddress{P.O. Box 1212}
	\city{Beijing} 
	\country{China} 
	%\postcode{100190}
}
\email{luochengleo@gmail.com}
\affiliation{%
	\institution{$^3$CAS Key Lab of Network Data Science and Technology, Institute of Computing Technology, \\ Chinese Academy of Sciences}
	%\streetaddress{P.O. Box 1212}
	\city{Beijing} 
	\country{China} 
	%\postcode{100190}
}
\email{guojiafeng@ict.ac.cn}
\fi

\begin{abstract}
Learning to rank with biased click data is a well-known challenge.
A variety of methods has been explored to debias click data for learning to rank such as click models, result interleaving and, more recently, the unbiased learning-to-rank framework based on inverse propensity weighting.
%In order to learn an unbiased ranker, IR researchers have explored a variety of methods to debias click data for learning to rank.
%Example studies include click models, online learning to rank and, more recently, the unbiased learning-to-rank framework that directly leverage click bias in the training of ranking models.
Despite their differences, most existing studies separate the estimation of click bias (namely the \textit{propensity model}) from the learning of ranking algorithms.
To estimate click propensities, they either conduct online result randomization, which can negatively affect the user experience, or offline parameter estimation, which has special requirements for click data and is optimized for objectives (e.g. click likelihood) that are not directly related to the ranking performance of the system. 
%is not adaptive to the change of user behaviors. 
In this work, we address those problems by unifying the learning of propensity models and ranking models.
We find that the problem of estimating a propensity model from click data is a dual problem of unbiased learning to rank.
Based on this observation, we propose a Dual Learning Algorithm (DLA) that jointly learns an unbiased ranker and an \textit{unbiased propensity model}.
% automatically conduct unbiased learning to rank with 
% learns propensity models and ranking models simultaneously.
DLA is an automatic unbiased learning-to-rank framework as it directly learns unbiased ranking models from biased click data without any preprocessing.
It can adapt to the change of bias distributions and is applicable to online learning.
Our empirical experiments with synthetic and real-world data show that the models trained with DLA significantly outperformed the unbiased learning-to-rank algorithms based on result randomization and the models trained with relevance signals extracted by click models.

%use online result randomization to estimate click propensities.
%require online result randomization experiments or offline click model to 

%Implicit user feedbacks from click data are critical resource for supervised learning to rank.
%Through they can be easily collected in large scale, they are   

\end{abstract}

\iffalse



\fi


\maketitle

%!TEX root=SIGIR18-NoisyLTR.tex
\section{Introduction}

% Clicks are important
%Machine learning techniques have become more and more popular in the industry and academe of Information Retrieval (IR)~\cite{liu2009learning}.
Machine learning techniques for Information Retrieval (IR) become widely used in both academic research and commercial search engines~\cite{liu2009learning}.
Although there have been studies that use unsupervised data or pseudo supervision for learning-to-rank models~\cite{ai2016analysis,dehghani2017neural}, the best retrieval system is typically constructed based on supervised learning.
Many of the state-of-the-art retrieval systems today make use of deep models~\cite{guo2016deep,Mitra:2017:LMU:3038912.3052579}, which require large amounts of labeled data.
%Most deep retrieval models~\cite{xxx} requires supervised training with millions of annotated query-document pairs. 
%With the increasing popularity of machine learning techniques in the industry and academic community of Information Retrieval (IR)~\cite{xxx}, modern retrieval systems are thirsty for annotated data.
Despite the development of crowdsourcing systems~\cite{kittur2008crowdsourcing,doan2011crowdsourcing}, obtaining large-scale and high quality human annotations (e.g. TREC-style relevance judgments) is still expensive, if not impossible.
%There are some studies that try to use unsupervised  pseudo supervision for the training of learning-to-rank models 
% for many ranking applications such as Web search, personal search and product recommendation.
Therefore, implicit feedback such as clicks are still the most attractive data source for the training of ranking systems.  

%thrist for data and data annotation
%despite efforts in xxx and cloud sourcing, obtaining large scale and high quality human annotations
%is still expensive, if not impossible, for many Ir appliations
%Therefore, implicit signals like clicks are still attractive   

% Clicks are biased
Directly training a ranking model to optimize click data, however, is infeasible because click data are heavily biased~\cite{joachims2005accurately,keane2006modeling,joachims2007evaluating,yue2010beyond}.
In particular, the order of documents in a search engine result page (SERP) has a strong influence on where users click~\cite{joachims2005accurately}.
Studies of position bias show that users tend to examine and click results on the top of a SERP while ignoring those on the bottom.
A naive method that treats click/non-click signals as positive/negative feedback will lead to a ranking model that optimizes the order of a search result page but not the relevance of documents.
%A simple corrected method that uses clicks as preferences between clicked and skipped documents is helpful but still problematic as it tends to reverse the order of documents~\cite{xxx}. 
%Simply treating clicks as preferences between clicked and skipped document 
%An irrelevant document could have a high click probability when placed in front of a relevant document.

% Click models, same query multiple times,
To leverage the full power of click data for learning to rank, IR researchers have attempted to debias click data before training ranking models.
One such effort is the development of click models.
The basic idea of click model is to make hypotheses about user browsing behaviors and estimate true relevance feedback by optimizing the likelihood of the observed user clicks.
Such methods work well on head queries in Web search but not on tail queries or other retrieval applications where multiple observations of same query-document pairs may not be available (e.g. personal search~\cite{wang2016learning}).
Also, the construction of click models is separated from the learning of ranking models. 
Click models are usually optimized for the likelihood of observed clicks but not the ranking performance of the overall system.
Their parameters need to be re-estimated whenever there are changes in user behaviors. %or new features introduced to search engines.
%We cannot conduct online learning to rank with click models because we need to infer the supervision signals offline and re-estimates the parameters whenever we introduce new features to search engines and there are changes in user behaviors.

Another effort to debias click data is result interleaving~\cite{yue2009interactively,chapelle2012large,raman2013learning,swaminathan2015batch,swaminathan2015counterfactual,schuth2016multileave}.
By collecting clicks on swapped results from the same result list, we can obtain unbiased pair preferences for documents and use them to train learning-to-rank models in an online manner.
This paradigm, however, introduces non-deterministic ranking functions into the product system~\cite{joachims2017unbiased}.
It may hurt the user experience by putting more irrelevant documents on the top of SERPs.
Based on these problems, a new research direction emerged recently that focuses on directly training ranking models with biased click data, which is often referred to as \textit{unbiased learning to rank}~\cite{wang2016learning,joachims2017unbiased}.  
The unbiased learning-to-rank framework treats click bias as a counterfactual effect and debiases user feedback by weighting each click with their Inverse Propensity Weights~\cite{rosenbaum1983central}.
It uses a propensity model to quantify click biases and does not explicitly estimate the query-document relevance with training data. 
As theoretically proven by Joachims et al.~\cite{joachims2017unbiased}, given the correct bias estimation, ranking models trained with click data under this framework will converge to the same model trained with true relevance signals.  

Despite their advantages, existing unbiased learning-to-rank algorithms share a common drawback with click models as they need a separate experiment to estimate click bias.
%, which makes them not adaptive for user behavior changes and not applicable for online learning to rank.
One of the most popular methods for click bias estimation is result randomization~\cite{wang2016learning,joachims2017unbiased}, which randomizes the order of documents so that the collected user clicks on randomized SERPs can reflect the examination bias of users on each result position.
%Although it can be conducted on a small proportion of search engine traffic separately, this paradigm is similar with result interleave as they both hurt user experience.
This paradigm is similar to result interleaving as they both can negatively affect user experience.
Additionally, because result randomization needs to be conducted on a proportion of search engine traffic separately, existing unbiased learning-to-rank models cannot adapt to changes in user behavior automatically. 
%Additionally, it is not adaptive to the changes of user behaviors and thus not suitable for online learning to rank.

%This paradigm, however, has two important limitations.
%First, randomized experiments are expensive.
%Although IR researchers have developed different strategies to reduce its cost~\cite{xxx}, randomizing search result order will inevitably hurt user experiences.
%This is intolerable for user-centric applications and small business.   
%Second, similar with click models, existing unbiased learning-to-rank algorithms cannot adapt to the change of user behaviors. 
%Users evolve rapidly when we introduce new features into modern search engines, which means that off-line estimations of click bias could be out-of-date frequently.
%Repeating the off-line experiment is not preferable especially when it is expensive. 

% idea
% advantages
% disadvantages
% randomize experiments
% not adaptive to change

% Cannor adapt to behavior changes

% Our model
% 
In this paper, we introduce a new framework for automatic unbiased learning to rank.
Most limitations of existing unbiased learning-to-rank models are caused by their additional user experiments for propensity estimation.
%As we can directly learn unbiased rankers from click data with the help of propensity models, we can also automatically estimate propensity models from raw clicks with the help of unbiased rankers.
%As position bias prevents us to estimate the true relevance of documents with raw clicks, the variance of document relevance in search sessions also prevents us to estimate position propensity directly from click data.
As Wang et al.~\cite{wang2016learning} and Joachims et al.~\cite{joachims2017unbiased} observed that unbiased rankers can be directly learned from user clicks with the help of propensity models, we observed that click propensity can be automatically estimated with click data given an unbiased ranking model.
%As Wang et al.~\cite{wang2016learning} and Joachims et al.~\cite{joachims2017unbiased} observed that a good propensity model can help us train an unbiased ranker, we observed that an unbiased ranker can also help us estimate an effective propensity model.
We formulate the problem of automatically estimating a propensity model from user clicks as \textit{unbiased propensity estimation} and propose a Dual Learning Algorithm (DLA) for unbiased learning to rank.
DLA jointly learns propensity models and ranking models based on raw click data.
Since it doesn't rely on any result randomization or offline experiments, DLA should be preferable in production systems and applicable to online learning to rank.
Furthermore, we theoretically prove that models trained with DLA will converge to their global optima under certain circumstances.
To evaluate the effectiveness of DLA in practice, we conducted both simulation and real-world experiments.
% to compare the performance of DLA with ranking models trained with click model signals and existing unbiased learning-to-rank framework.
Empirical experimental results show that models trained with DLA are adaptive to changes in user behavior and significantly outperformed the models trained with click model signals and existing unbiased learning-to-rank frameworks. 
%that jointly learns propensity models and ranking models from raw user clicks.
%DLA 
%DLA doesn't rely on any result randomization or offline experiment, so it is applicable 

%In this paper, we introduce a new framework for unbiased learning to rank.
%Motivated by the problems outlined above, our goal is to automate the learning process of an unbiased ranker without using any off-line experiments.
%To do that, we develop a dual learning algorithm.
%The dual learning algorithm can simultaneously learn an unbiased ranker with an unbiased bias estimation model.
%As a good bias estimation can help us train an unbiased learning-to-rank model, we argue that a good ranking model can also help us estimate an unbiased bias model.
%With the dual learning algorithms, we jointly learn the ranker and the bias model with click data. 
%Different from existing models, our framework is an end-to-end method that requires no off-line experiments and is fully automatic.
%It can be directly applied to any ranking problem with implicit feedback and also can be used in online manner. 

% Our contribution
The contributions of this paper are summarized as follows:
\begin{itemize}
\item We formulate a problem of \textit{unbiased propensity estimation} and discuss its relationship with unbiased learning to rank.
\item We propose a Dual Learning Algorithm that automatically and jointly learns unbiased propensity models and ranking models from raw click data.
\item We conduct both theoretical analysis and empirical experiments to understand the effect of the joint learning process used by DLA.
\end{itemize}

\iftrue
% Section organization
The rest of the paper is organized as follows.
In Section~\ref{sec:related}, we review previous work on learning to rank with click data.
%Then we describe the existing unbiased learning-to-rank frameworks in Section~\ref{sec:ULTR}.
%We formally introduce DLA in Section~\ref{sec:our_approach} and discuss our experiment setup and results in Section~\ref{sec:setup} and Section~\ref{sec:results}.
We introduce existing unbiased learning-to-rank frameworks and the Dual Learning Algorithm in Section~\ref{sec:ULTR}\&\ref{sec:our_approach}.
Our experiment setup and results are described in Section~\ref{sec:setup}\&\ref{sec:results}.
Finally, we conclude this paper and discuss future work in Section~\ref{sec:conclusion}.
\fi

%!TEX root=SIGIR18-NoisyLTR.tex
\section{Related Work}\label{sec:related}

%Learning to rank with click data has been extensively studied in IR, especially in Web search.
There are two groups of approaches to handle biased user feedback for learning to rank. 
The first group focuses on debiasing user clicks and extracting reliable relevance signals. 
%The second group tries to design algorithms that directly learn unbiased rankers from biased feedback.
The second group tries to directly learn unbiased rankers from biased feedback.

\textbf{Debias User Feedback from Click Data}.
As shown by prior work~\cite{joachims2005accurately,keane2006modeling,joachims2007evaluating,yue2010beyond,wang2013incorporating}, implicit user feedback from click data is severely biased. 
For the robustness of learning-to-rank models trained with click data, IR researchers have conducted extensive studies on user browsing behaviors and constructed click models~\cite{craswell2008experimental,dupret2008user,chapelle2009dynamic,xu2010temporal,wang2013content,wang2013incorporating} to extract real relevance feedback from click signals.
For example, Craswell et al.~\cite{craswell2008experimental} designed a Cascade model to separate click bias from relevance signals by assuming that users read search result pages sequentially from top to bottom.
Dupret and Piwowarski~\cite{dupret2008user} proposed a User Browsing Model (UBM) that allows users to skip some results by computing the examination probability of documents according to their positions and last clicks.
To further incorporate search abandon behaviors, Chapelle and Zhang~\cite{chapelle2009dynamic} constructed a Dynamic Bayesian Network model (DBN) that uses a separate variable to model whether a user is satisfied by a click and ends the search session. 
In spite of their underlying hypotheses, the goal of click models is to estimate the ``true" relevance feedback and use them for learning to rank.
Most click models need to be constructed offline and require each query-document pair to appear multiple times for reliable relevance estimation.
In contrast, we propose to estimate click bias automatically and jointly with the learning of ranking systems so that our model can be applied to online learning and retrieval applications where multiple appearances of query-document pairs are not available (e.g. email search).

To avoid the modeling of user behaviors, another direction of research tries to collect unbiased relevance feedback directly from users.
For instance, it has been shown that presenting randomized ranked lists and collecting user clicks accordingly can reveal the true preferences between swapped documents~\cite{yue2009interactively,chapelle2012large,raman2013learning,swaminathan2015batch,swaminathan2015counterfactual,schuth2016multileave}.
Yue and Joachims~\cite{yue2009interactively} used result interleaving to collect reliable user feedback and used Dual Bandit Gradient Descent (DBCG) to learn ranking models in an online manner.
Schuth et al.~\cite{schuth2016multileave} extended DBCG and proposed Multileave Gradient Descent to compare multiple rankings and find good rankers. 
The effectiveness of these learning paradigms has been justified in theory~\cite{yue2009interactively}, but they are not popular in practice because result interleaving hurts ranking quality and introduces non-deterministic factors into search engines~\cite{joachims2017unbiased}.
The approach proposed in our work is also applicable to online learning to rank but it learns rankers from real user clicks without any result randomization.

%debias user feedback from click data
%Two way, make assumption to extract and remove from the beginning
%click model, Cascade, UBM, DBN, pscm
%remove from begining, iteractive, multo, xxx
%similar and difference: offline training, interleaves, joint leaning

\textbf{Unbiased Learning to Rank}.
%Unbiased learning to rank is a new approach to account click bias for learning-to-rank algorithms proposed recently~\cite{wang2016learning,joachims2017unbiased}.
Based on the intrinsic limitations of click models, IR researchers have explored a new approach to account for click bias for learning to rank.
Instead of inferring relevance signals by optimizing observed click likelihood~\cite{craswell2008experimental,dupret2008user,chapelle2009dynamic,xu2010temporal,wang2013content,wang2013incorporating}, the examination propensity of each document can be estimated through result randomization and used to construct unbiased ranking loss for learning to rank.
For example, Wang et al.~\cite{wang2016learning} proposed estimating the selection bias at query level through a randomization experiment and used Inverse Propensity Weighting (IPW)~\cite{rosenbaum1983central} to debias the training loss computed with click signals.
Joachims et al.~\cite{joachims2017unbiased} analyzed the IPW framework for unbiased learning to rank and showed that it can find the unbiased ranker theoretically and empirically.
The framework of existing unbiased learning-to-rank algorithms doesn't require multiple observations for each query-document pair.
In contrast to online learning algorithms with result interleaving, it constructs a deterministic ranking model~\cite{joachims2017unbiased}.
Nonetheless, most existing unbiased learning-to-rank algorithms still rely on a separate result randominzation experiment to estimate the propensity model for IPW.
They are not immune to the problems resulting from result randomization.
Wang et al.~\cite{wang2018position} proposed a novel EM algorithm to estimate click propensity without result randomization, but models trained with their or other unbiased learning-to-rank framework are not adaptive.
The EM process and the result randomization have to be conducted every time when there is any change in search engine interfaces or user behaviors. 
%Also, models trained with existing unbiased learning-to-rank framework are not adaptive because result randomization has to be conducted every time when there is any change in search engine interfaces or user behaviors. 
Similar to previous studies~\cite{wang2016learning,joachims2017unbiased,wang2018position}, we adopt the IPW framework for unbiased learning to rank. %with those used by Wang et al.~\cite{wang2016learning} and Joachims et al.~\cite{joachims2017unbiased} for unbiased learning to rank.
However, we discard result randomization and automate the entire unbiased learning-to-rank framework so that propensity models and unbiased rankers can be jointly learned with raw user clicks.  

%unbiased learning to rank
%Wang 
%Joachism
%IPW The Central Role of the Propensity Score in Observational Studies for Causal Effect
%similarity difference, 

%interpret bias
%reduce bias feedback from the begining
	%Interactively optimizing information retrieval systems as a dueling bandits problem
	%Multileave Gradient Descent for Fast Online Learning to Rank
	%A. Swaminathan and T. Joachims. Batch learning from logged bandit feedback through counterfactual risk minimization
	%Counterfactual Risk Minimization: Learning from Logged Bandit Feedback
		%-> show interleaving results to users to collect unbiased feedback

%!TEX root=SIGIR18-NoisyLTR.tex

\begin{table}[t]
	\setlength{\belowcaptionskip}{-10pt}
	\caption{A summary of notations used in this paper.}
	\small
	\def\arraystretch{1.15}%  1 is the default, change whatever you need
	\begin{tabular}
		{| p{0.09\textwidth} | p{0.35\textwidth}|} \hline
		$\mathcal{Q}, Q, q$ & The universal set of queries $\mathcal{Q}$, a sample set $Q$ and a query instance $q \sim P(q)$.   \\\hline
		$S$, $\theta$, $E$, $\phi$ & A ranking system $S$ parameterized by $\theta$ and a propensity model $E$ parameterized by $\phi$. \\\hline
		%$\mathcal{E}, $ & The universal set of examination models $\mathcal{E}$ and an arbitrary instance $E$.\\\hline
		$\pi_q$, $x$, $i$, $y$ & A ranked list $\pi_q$ produced by $S$ for $q$, a document $x$ on the $i$th position in $\pi_q$ and its relevance $y$.\\\hline
		$\mathbf{o}_q$, $\mathbf{r}_q$, $\mathbf{c}_q$ & The sets of Bernoulli variables that represent whether a document $x$ in $q$ is observed ($o_q^x$), perceived as relevant ($r_q^x$) and clicked ($c_q^x$).\\\hline
		%$f_q^T(x)$, $f_q^S(x, \theta)$ & The true probability of $P(r_q^x=1|x,q) = f_q^T(x)$ and the estimated probability $P_S(r_q^x=1|x,q) = f_q^S(x, \theta)$ computed by the ranking system $S$ parameterized by $\theta$. \\\hline
		%$g_q^T(x)$, $g_q^E(x, \phi)$ & The true probability of $P(o_q^x=1|x,q,S_q) = g_q^T(x)$ and the estimated probability $P_E(o_q^x=1|x,q,S_q) = g_q^E(x, \phi)$ computed by the examination model $E$ parameterized by $\phi$. \\\hline	
		%$\mathcal{L}(S)$, $l(S, q) $ & The overall loss of system $S$ ($\mathcal{L}(S)$) and the local loss of $S$ on $q$ ($l(S, q)$). \\\hline
		%$\alpha$ & The learning rate $\alpha$ used in model training.\\\hline
	\end{tabular}\label{tab:notation}
	%\vspace{-10pt}
\end{table}

\section{Unbiased Learning to Rank}\label{sec:ULTR}
In this section, we discuss the existing unbiased learning-to-rank framework.
The core of unbiased learning to rank is to debias loss functions built with user clicks so that the ranking model converges to the model trained with true relevance labels.

A summary of notations used in this paper is shown in Table~\ref{tab:notation}.
Without the loss of generality, we describe learning to rank with true relevance information as follows.
Let $\mathcal{Q}$ be the universal set of all possible queries and $q$ be an instance from $\mathcal{Q}$ which follows the distribution of $q \sim P(q)$.
Suppose that we have a ranking system $S$ and a loss function $l$ defined over $S$ and $q$, then the global loss $\mathcal{L}$ of $S$ is defined as
$$
\mathcal{L}(S) = \int_{q\in\mathcal{Q}} \!\!l(S,q)~dP(q)
$$
The goal of learning to rank is to find the best ranking system $S$ that minimizes $\mathcal{L}(S)$. 
Because $\mathcal{L}(S)$ cannot be computed directly, we often estimate it empirically based on a separate training query set $Q$ and the uniform assumption on $P(q)$:
$$
\hat{\mathcal{L}}(S) = \frac{1}{|Q|}\sum_{q\in Q} l(S,q)
$$
%Therefore, the optimization goal of a typical learning-to-rank algorithms is to find the best $S$ that minimize the empirical loss $\hat{\mathcal{L}}(S)$ from the universal set of ranking systems $\mathcal{S}$~\cite{empirical risk minimization}:
%$$
%\hat{S} = \text{argmin}_{S\in \mathcal{S}} \big\{\hat{\mathcal{L}}(S)\big\}
%$$

%The key for estimating $\hat{\mathcal{L}}(S)$ is the computation of the local ranking loss $l(S,q)$.
Usually, $l(S,q)$ is defined over the order of documents and their relevance with the query. 
Let $\pi_q$ be the ranked list retrieved by $S$ for query $q$, $x$ be a document in $\pi_q$ and $y$ be the binary label that denotes whether $x$ is relevant to $q$.
In most cases, we are only concerned with the position of relevant documents ($y=1$) in retrieval evaluations (e.g. MAP, nDCG~\cite{jarvelin2002cumulated}, ERR~\cite{chapelle2009expected}), so we can formulate the local ranking loss $l(S,q)$ as:
\begin{equation}
\begin{split}
l(S,q) = \sum_{x \in \pi_q, y=1} \!\!\Delta(x, y|\pi_q)
\end{split}
\label{equ:rank_loss}
\end{equation}
where $\Delta(x, y|\pi_q)$ is a function that computes the individual loss on each relevant document in $\pi_q$. 
%Examples for $l(S,q)$ and $\Delta(x, y|\pi_q)$ includes MAP and precision, nDCG and DCG~\cite{xxx}, reciprocal rank and ERR~\cite{xxx} etc.

The relevance labels $y$ are typically elicited either explicitly through expert judgments or implicitly via user feedback.
The former is often considered to be more reliable, but it is expensive or impossible to obtain in many retrieval scenarios (e.g. personal search).
Also, it generates relevance judgments based on the aggregation of all intents that underlie the same query string with the distributions estimated by judges but not real users.
The latter, which refers to clicks collected from real users, are cheap yet heavily biased.
It is affected by multiple factors including presentation bias~\cite{yue2010beyond}, trust bias~\cite{keane2006modeling,joachims2007evaluating} and, most commonly, position bias~\cite{joachims2005accurately}.  
To utilize the relevance information hidden in user clicks, we must debias the click signals before applying it to learning-to-rank frameworks.
One of the standard methods for bias correction is the inverse propensity weighting algorithm~\cite{wang2016learning,joachims2017unbiased}.

\subsection{Inverse Propensity Weighting}\label{sec:IPW}

Inverse propensity weighting (IPW) is first proposed for unbiased learning to rank by Wang et al.~\cite{wang2016learning} and Joachims et al.~\cite{joachims2017unbiased}.
It introduces a counterfactual model that removes the effect of click bias.
%Similarly with previous studies, we focus on the discussion of position bias removal with IPW in this paper. 
%It is worth noticing that one can easily extend the model to incorporate other biases~\cite{xxx}.
Let $\mathbf{o}_q, \mathbf{c}_q$ be the sets of Bernoulli variables that represent whether the documents in $\pi_q$ are observed and clicked by a user.
For simplicity, we assume that $x$ will be clicked ($c_q^x=1$) when it is observed ($o_q^x=1$) and relevant ($y=1$).
The main idea of IPW is to optimize ranking systems $S$ with an inverse propensity weighted loss defined as  
\begin{equation}
\begin{split}
l_{IPW}(S,q) &= \!\!\sum_{x \in \pi_q} \!\Delta_{IPW}(x, y|\pi_q) = \!\!\!\!\!\!\!\!\sum_{\substack{x \in \pi_q, o_q^x =1, y=1}} \!\!\!\!\frac{\Delta(x, y|\pi_q)}{P(o_q^x =1| \pi_q)}
\end{split}
\label{equ:rank_IPW}
\end{equation}

There are two important properties of the inverse propensity weighted loss.
First, $\Delta_{IPW}(x, y|\pi_q)$ is computed only when $x$ is both observed and relevant, so we can ignore non-clicked documents in $l_{IPW}(S,q)$.
This is essential because we do not know the reason that causes $c_q^x=0$ (either $o_q^x=0$ or $y=0$ or both).
% the behavior of not clicking is often noisy
Second, the IPW loss is theoretically principled because it is an unbiased estimation of $l(S,q)$.
As shown by Joachims et al.~\cite{joachims2017unbiased}, the expectation of the inverse propensity weighted loss is
\begin{equation}
	\begin{split}
		\mathbb{E}_{\mathbf{o}_q}[l_{IPW}(S,q)] &= \mathbb{E}_{\mathbf{o}_q}\big[\!\!\!\!\sum_{x \in \pi_q, o_q^x =1,y=1} \!\!\!\!\frac{\Delta(x, y|\pi_q)}{P(o_q^x =1| \pi_q)}\big] \\
		&= \mathbb{E}_{\mathbf{o}_q}\big[\sum_{x \in \pi_q,y=1} \frac{o_q^x \cdot \Delta(x, y|\pi_q)}{P(o_q^x =1| \pi_q)}\big] \\
		%&= \sum_{x \in \pi_q} \mathbb{E}_{\mathbf{o}_q}\big[\frac{o_q^x \cdot \Delta(x, y|\pi_q)}{P(o_q^x =1| \pi_q)}\big]  \\
		&= \sum_{x \in \pi_q,y=1} \mathbb{E}_{\mathbf{o}_q}\big[o_q^x\big] \cdot \frac{\Delta(x, y|\pi_q)}{P(o_q^x =1| \pi_q)} \\
		&= \sum_{x \in \pi_q,y=1} P(o_q^x =1| \pi_q) \cdot \frac{\Delta(x, y|\pi_q)}{P(o_q^x =1| \pi_q)} \\
		&= \!\!\!\!\!\sum_{x \in \pi_q, y=1}\!\!\!\! \Delta(x, y|\pi_q) = l(S,q)
	\end{split}
	\label{equ:ipw_expectation}
\end{equation}
The ranking model trained with clicks and the IPW loss will converge to the same model trained with true relevance labels.
Thus, the whole learning-to-rank framework is unbiased. 

\subsection{Randomization-based Estimation}~\label{sec:randomization}
The key of the IPW algorithm is the estimation of propensity model $P(o_q^x =1| \pi_q)$. 
%The propensity of users to observe a document depends on multiple factors~\cite{xxx}.
Although the framework can be easily extended to other biases~\cite{joachims2017unbiased}, most existing work on unbiased learning to rank only focuses on the effect of position bias~\cite{joachims2005accurately} for simplicity.
This work assumes that $P(o_q^x =1| \pi_q)$ only depends on the position of $x$:
$$
P(o_q^x =1| \pi_q) = P(o_i=1)
$$
where $i$ is the position of $x$ in the ranked list $\pi_q$.

A simple yet effective solution to estimate a position-based propensity model is result randomization~\cite{wang2016learning,joachims2005accurately}.
The idea of result randomization is to shuffle the order of documents and collect user clicks on different positions to compute the propensity model.
Because the expected document relevance is the same on all positions, it is easy to prove that result randomization method produces an optimal estimator for position-based propensity model:
\begin{equation}
\begin{split}
\mathbb{E}[c_k] &= \int_{(q,x,\pi_q), i=k}\!\!\!\!\!\!P(c_q^x=1|\pi_q)~dP(q,x,\pi_q) \\
&= \int_{(q,x,\pi_q), i=k}\!\!\!\!\!\!P(o_i=1) \cdot y ~ dP(q,x,\pi_q) \\
&= P(o_k=1) \cdot \int_{(q,x,\pi_q), i=k} \!\!\!\!\!\!y ~ dP(q,x,\pi_q) \\
&\propto P(o_k=1) 
\end{split}
\end{equation}

Despite its simplicity and effectiveness, result randomization has intrinsic drawbacks that limit its applications.
First, it can significantly affect the user experience.
Shuffling documents in the ranked list will inevitably put more irrelevant results in high positions.
Previous studies have explored several strategies (e.g. pair-based randomization~\cite{joachims2005accurately}) to alleviate this problem, but none of them can solve it completely.
Second, the use of result randomization makes existing unbiased learning-to-rank framework less efficient and adaptive. 
%The results of randomization experiments could be out-of-date in practice.
Randomization experiments is time-consuming and has to be conducted separately with the training of ranking models. 
It is painful to re-train the whole system and thus difficult to keep the model updated with changes in user behaviors.
As far as we know, most existing unbiased learning-to-rank algorithms rely on result randomization to estimate propensity model, which makes them vulnerable to the two problems discussed above.
To solve them, we discard randomization experiments completely and propose to automatically learn both the ranking model and the propensity model based on user clicks.

%not solvable, to use with clicks, we propose xxx
%  painful for model re-training.
%However, the need to update ranking models is common because user behavior could change when we introduce new features into the interface of our retrieval systems.

%To solve the problems of existing unbiased learning-to-rank algorithms based on , 

%Because the randominzation experiment
%This makes the training of ranking models under existing unbiased learning-to-rank frameworks expensive and time consuming.

%Because the propensity estimator based on result randomization cannot take advantage of the    
%the use of result randomizations makes existing unbiased learning-to-rank frameworks not adaptive to the change of user behavior.

%As result randomization requires a separate experiment to collect user clicks, its results will be out-of-date whenever user behavior changes.     
% have to be conducted separately the training of ranking models.

%Such a two-step process used by existing unbiased learning-to-rank algorithm requires us to 
%The framework of existing unbiased learning-to-rank algorithms requires 
%The common methodology for unbiased learning to rank in practice is to estimate the propensity model based on the randomization experiments on a small proportion of search engine traffics and        
%There are other propensity estimation methods

\section{Our Approach}~\label{sec:our_approach}

We now describe our approach for automatic unbiased learning to rank.
The key component that limits existing unbiased learning-to-rank algorithms to be fully automatic is the estimation of click propensity.
In this work, we find that estimating a propensity model with user clicks is actually a dual problem of unbiased learning to rank, which could be unbiasedly learned in a similar way.
%Inspired by the definition of unbiased ranker, we formalize the problem of learning the true position propensity as \textit{unbiased propensity estimation}.
Thus, we refer to it as \textit{unbiased propensity estimation} and propose a Dual Learning Algorithm that jointly learns the unbiased ranking model and the propensity model with click data.
Theoretical analysis shows that our approach is guaranteed to find the unbiased ranker and propensity model under certain circumstances.

\subsection{Unbiased Propensity Estimation}\label{sec:IRW}
%The key component that limits existing unbiased learning-to-rank algorithms to be fully automatic is the estimation of propensity. 
Let $\mathbf{r}_q$ be a set of Bernoulli variables which denote whether the documents in $\pi_q$ will be perceived as relevant when users observe them. 
Then the probability that a document $x\in\pi_q$ will be clicked can be computed as
$$
P(c_q^x =1 | \pi_q) = P(o_q^x=1 | \pi_q) \cdot P(r_q^x=1 | \pi_q)
$$  
Because users click a search result ($c_q^x=1$) only when it is both observed ($o_q^x=1$) and perceived as relevant ($r_q^x=1$), we cannot directly infer the relevance of a document without knowing whether it has been examined.
Similarly, we cannot estimate the propensity of examination without knowing whether the documents are relevant or not.
From this point of view, the problem of estimating examination propensity is symmetric with the problem of estimating real document relevance from user clicks.
Since the latter is referred to as \textit{unbiased learning to rank}, we formulate the former problem as \textit{unbiased propensity estimation}. 
%As previous studies formulate the latter problem as unbiased learning to rank, we formulate the former problem as \textit{unbiased propensity estimation}. 

We now formally describe the problem of unbiased propensity estimation. 
Similar to learning to rank, the goal of propensity estimation is to find the optimal propensity model $E$ that minimizes a global loss function:
$$
\mathcal{L}(E) = \int_{q\in\mathcal{Q}} \!\!\!\!\!\!\! l(E,q)~dP(q)
$$
where $l(E,q)$ is a function that computes the local loss of $E$ in query $q$.
Suppose that we only care about the performance of propensity estimation on documents that have been observed by users, then $l(E,q)$ can be defined as
\begin{equation}
\begin{split}
l(E,q) = \!\!\!\!\sum_{x \in \pi_q, o_q^x=1} \!\!\!\!\Delta(x, o_q^x|\pi_q)
\end{split}
\label{equ:exam_loss}
\end{equation}

Under this formulation, it is obvious that the learning of a propensity model is similar to the learning of a ranking function.
Thus, the inverse propensity weighting algorithm for unbiased learning to rank can also be directly applied to unbiased propensity estimation.
Similar to Equation~(\ref{equ:rank_IPW}), we define the \textit{Inverse Relevance Weighted} (IRW) loss for $E$ as 
\begin{equation}
\begin{split}
l_{IRW}(E,q) &= \sum_{x \in \pi_q} \Delta_{IRW}(x, o_q^x|\pi_q) = \!\!\!\!\!\!\!\!\!\!\!\!\sum_{x \in \pi_q, o_q^x =1, r_q^x=1} \!\!\!\!\!\!\frac{\Delta(x, o_q^x|\pi_q)}{P(r_q^x =1| \pi_q)}
\end{split}
\label{equ:exam_IPW}
\end{equation}
Following the same logic flow in Equation~(\ref{equ:ipw_expectation}), we can easily prove that this is an unbiased estimate of $l(E,q)$ because:
\begin{equation}
\begin{split}
\mathbb{E}_{\mathbf{r}_q}[l_{IRW}(E,q)] &= \mathbb{E}_{\mathbf{r}_q}\big[\!\!\!\!\sum_{x \in \pi_q, o_q^x=1, r_q^x =1} \!\!\!\!\frac{\Delta(x, o_q^x|\pi_q)}{P(r_q^x =1|\pi_q)}\big] \\
&= \mathbb{E}_{\mathbf{r}_q}\big[\sum_{x \in \pi_q, o_q^x=1} \frac{r_q^x \cdot \Delta(x, o_q^x|\pi_q)}{P(r_q^x =1| \pi_q)}\big] \\
%&= \sum_{x \in \pi_q} \mathbb{E}_{\mathbf{r}_q}\big[\frac{r_q^x \cdot \Delta(x, o_q^x|\pi_q)}{P(r_q^x =1| y)}\big]  \\
&= \sum_{x \in \pi_q, o_q^x=1} \mathbb{E}_{\mathbf{r}_q}\big[r_q^x\big] \cdot \frac{\Delta(x, o_q^x|\pi_q)}{P(r_q^x =1| \pi_q)} \\
&= \sum_{x \in \pi_q, o_q^x=1} P(r_q^x =1| \pi_q) \cdot \frac{\Delta(x, o_q^x|\pi_q)}{P(r_q^x =1| \pi_q)} \\
&= \!\!\!\!\!\sum_{x \in \pi_q, o_q^x=1} \!\!\!\!\Delta(x, o_q^x|\pi_q) = l(E,q)
\end{split}
\label{equ:exam_expection}
\end{equation}
%Therefore, we can directly learn the propensity model from click data given a model of $P(r_q^x=1|\pi_q)$.

If we compare the problem of unbiased learning to rank with the problem of unbiased propensity estimation, we can see that the goal of the former is to estimate $P(r_q^x=1|\pi_q)$ while the goal of the latter is to estimate $P(o_q^x=1|\pi_q)$.
This indicates that a good ranking model can help us estimate a good propensity model and  vice versa.
Based on this observation, we propose a Dual Learning Algorithm to jointly learn the propensity model and the ranking model with click data.

%The learning of a propensity model based on $l_{IPW}(E,q)$

%IPW

\subsection{Dual Learning Algorithm}

\iffalse
\begin{algorithm}[t]
	\caption{Dual Learning Algorithm}
	\label{alg:DLA}
	\KwIn{$Q=\{q, \pi_q, \mathbf{c}_q\}, \alpha$}
	\KwOut{$\theta, f, \phi, g$}
	\nl Initialize $\theta \Leftarrow 0$, $\phi \Leftarrow 0$ \\
	\nl \Repeat{Convergence}{
		\nl Randomly sample a batch $(q, \pi_q, \mathbf{c}_q)$ from $Q$
		\For{each batch of $(q, \pi_q, \mathbf{c}_q) \subseteq Q$}{
			%\For{$x \in \pi_q$}{Compute $f_q^x(\theta)$, $g_q^x(\phi)$.}
			\nl \For{$x \in \pi_q$}{\nl Compute $P(r_q^x=1|\pi_q)$, $P(o_q^x=1|\pi_q)$ with Eq~\ref{equ:softmax}.}
		\nl Compute $\hat{L}(S,q)$, $\hat{L}(E,q)$ with Eq~\ref{equ:DLA_loss}. \\
		\nl $\theta = \theta + \alpha \cdot \frac{\partial \hat{L}(S,q)}{\partial \theta}$, $\phi = \phi + \alpha \cdot \frac{\partial \hat{L}(E,q)}{\partial \phi}$
		} 
	}
	\nl \Return $\theta, f, \phi, g$
\end{algorithm}
\fi

\begin{algorithm}[t]
	\small
	\caption{Dual Learning Algorithm}
	\label{alg:DLA}
	\KwIn{$Q=\{q, \pi_q, \mathbf{c}_q\}$, $f$, $g$, learning rate $\alpha$}
	\KwOut{$\theta, \phi$}
	\nl Initialize $\theta \Leftarrow 0$, $\phi \Leftarrow 0$ \\
	\nl \Repeat{Convergence}{
		\nl Randomly sample a batch $Q'$ from $Q$ \\
		\For{$(q, \pi_q, \mathbf{c}_q) \in Q'$}{
			%\For{$x \in \pi_q$}{Compute $f_q^x(\theta)$, $g_q^x(\phi)$.}
			\nl \For{$x \in \pi_q$}{\nl Compute $P(r_q^x=1|\pi_q)$, $P(o_q^x=1|\pi_q)$ with Eq~\ref{equ:softmax}.}
		} 
		\nl Compute $\hat{L}(S,q)$, $\hat{L}(E,q)$ with Eq~\ref{equ:DLA_loss}. \\
		\nl $\theta = \theta + \alpha \cdot \frac{\partial \hat{L}(S,q)}{\partial \theta}$, $\phi = \phi + \alpha \cdot \frac{\partial \hat{L}(E,q)}{\partial \phi}$
	}
	\nl \Return $\theta, \phi$
\end{algorithm}

% probability and loss function
The idea of the Dual Learning Algorithm (DLA) is to solve the problems of unbiased learning to rank and unbiased propensity estimation simultaneously.
It has two important components: the loss function $l(S,q)$, $l(E,q)$ and the estimation of $P(o_q^x=1|\pi_q)$ and $P(r_q^x=1|\pi_q)$.
Let $g_q^x(\phi)$ and $f_q^x(\theta)$ be the propensity score and ranking score produced by the propensity model $E$ (parameterized by $\phi$) and ranking model $S$ (parameterized by $\theta$) for document $x$ in query $q$.
%An overview of the DLA is shown in Algorithm~\ref{alg:DLA}. 
%Here we first describe the components of the DLA and then provide theoretical analysis on how it converge to the unbiased ranker and propensity model.
%\textbf{Algorithm Overview}.
%The DLA has two important components: the loss function $l(S,q)$, $l(E,q)$ and the propensity estimation of $P(r_q^x=1|\pi_q)$, $P(o_q^x=1|\pi_q)$.
As discussed previously, $l(S,q)$ and $l(E,q)$ only have values on clicked documents and click behavior only happens on documents that are observed and relevant. 
Thus, pointwise loss functions are likely to fail in the IPW framework because we only use relevant documents to train the ranking model.
%As discuss previously, $l(S,q)$ and $l(E,q)$ only have value on clicked documents, which fails point-wise loss functions because there are no negative feedback signal.
%In this work, we adapted a list-wise loss based on softmax-based cross entropy for the DLA:
Inspired by ~\cite{ai2018DLCM}, we adapted a list-wise loss based on softmax-based cross entropy for DLA as:
%Let $g_q^x$ and $f_q^x$ be the propensity score and ranking score produced by the propensity model $E$ and ranking model $S$ for document $x$ in query $q$. 
%Then we define:
\begin{equation}
\begin{split}
l(E,q) &= \!\!\!\!\!\!\!\!\!\!\!\!\sum_{x \in \pi_q, o_q^x=1, r_q^x=1}\!\!\!\!\!\!\!\!\!\!\!\!\Delta(x, o_q^x|\pi_q) = -\!\!\!\!\!\!\!\!\sum_{x \in \pi_q, c_q^x=1} \!\!\!\!\!\!\log\frac{e^{g_q^x(\phi)}}{\sum_{z \in \pi_q}e^{g_q^z(\phi)}} \\
l(S,q) &= \!\!\!\!\!\!\!\!\!\!\!\!\sum_{x \in \pi_q, o_q^x=1, r_q^x=1} \!\!\!\!\!\!\!\!\!\!\!\!\Delta(x, r_q^x|\pi_q) = -\!\!\!\!\!\!\!\!\sum_{x \in \pi_q, c_q^x=1} \!\!\!\!\!\!\log\frac{e^{f_q^x}(\theta)}{\sum_{z \in \pi_q}e^{f_q^z(\theta)}}
\end{split}
\label{equ:softmax_loss}
\end{equation}
The softmax-based cross entropy naturally converts the outputs of ranking models and propensity models into probability distributions, which are then used for the propensity and relevance estimation:
\begin{equation}
P_E(o_q^x\!=\!1|\pi_q) \!= \!\frac{e^{g_q^x(\phi)}}{\sum_{z \in \pi_q}\!\!e^{g_q^z(\phi)}}, ~~~~~~ P_S(r_q^x\!=\!1|\pi_q) \!=\! \frac{e^{f_q^x(\theta)}}{\sum_{z \in \pi_q}\!\!e^{f_q^z(\theta)}}
\label{equ:softmax}
\end{equation}
Note that other loss functions can also be adopted in DLA as long as they follow a similar probability framework in Equation~(\ref{equ:softmax_loss})\&(\ref{equ:softmax}).
We leave the investigation of other loss functions for future studies.

As shown in Equation~(\ref{equ:softmax}), the use of the softmax function assumes that the examination probabilities on different positions in a ranked list will sum up to 1, which is not true in practice.
This, however, does not hurt the effectiveness of model training. 
The predicted values for $P(r_q^x=1|\pi_q)$ and $P(o_q^x=1|\pi_q)$ have a minor effect on the unbiased learning process as long as their relative proportions are correct.
In fact, the actual inverse propensity weighted loss functions used in the DLA are:
\begin{equation}
\begin{split}
\hat{l}_{IRW}(E,q) &= -\!\!\!\!\!\!\!\!\sum_{x \in \pi_q, c_q^x=1}\!\! \frac{P_S(r_q^1\!=\!1|\pi_q)}{P_S(r_q^x\!=\!1|\pi_q)}\cdot\log\frac{e^{g_q^x(\phi)}}{\sum_{z \in \pi_q}\!\!e^{g_q^z(\phi)}}\\
\hat{l}_{IPW}(S,q) &= -\!\!\!\!\!\!\!\!\sum_{x \in \pi_q,c_q^x=1}\!\! \frac{P_E(o_q^1\!=\!1|\pi_q)}{P_E(o_q^x\!=\!1|\pi_q)} \cdot\log\frac{e^{f_q^x(\theta)}}{\sum_{z \in \pi_q}\!\!e^{f_q^z(\theta)}} 
\end{split}
\label{equ:IPW_softmax_loss}
\end{equation}
where $P(o_q^1=1|\pi_q)$ and $P(r_q^1=1|\pi_q)$ are the marginal probabilities for the first document in $\pi_q$.
%Because $P(o_q^0=1|\pi_q)$ and $P(r_q^0=1|\pi_q)$ are constants, 
The expected values of $\hat{l}_{IRW}(E,q)$ and $\hat{l}_{IPW}(S,q)$ are proportional to $l(S,q)$, $l(E,q)$, which doesn't affect the effectiveness of unbiased learning discussed in Equation~\ref{equ:ipw_expectation}\&\ref{equ:exam_expection}.
%We will further prove the effectiveness of the DLA in the following of this section.
Finally, the empirical loss of $S$ and $E$ can be computed as:
\begin{equation}
\hat{\mathcal{L}}(E) = \frac{1}{|Q|}\sum_{q\in Q} \hat{l}_{IRW}(E,q), \hat{\mathcal{L}}(S) = \frac{1}{|Q|}\sum_{q\in Q} \hat{l}_{IPW}(S,q)
\label{equ:DLA_loss}
\end{equation}
To compute the loss on a batch of queries $Q' \subseteq Q$, we can simply replace $Q$ with $Q'$ in Equation~(\ref{equ:DLA_loss}).

An overview of the complete algorithm is shown in Algorithm~\ref{alg:DLA}.
In DLA, we first initialize all parameters to zero. 
Then for each batch of queries, we compute $P(r_q^x=1|\pi_q)$, $P(o_q^x=1|\pi_q)$ with Equation~(\ref{equ:softmax}) and $\hat{L}(S,q)$, $\hat{L}(E,q)$ with Equation~(\ref{equ:DLA_loss}).
We update $\theta$ and $\phi$ with the derivatives of $\hat{L}(S,q)$ and $\hat{L}(E,q)$ respectively and repeat the process until the algorithm converges.

%This is an important property as it avoids the use of unclicked behavior

\subsection{Convergence Analysis}\label{sec:convergence}

% what we want to do here
%Previous studies on unbiased learning to rank and our discussion on unbiased propensity estimation prove that we can learn the optimal ranker given the correct propensity model and learn the optimal propensity model given the correct ranker.
As discussed in Section~\ref{sec:IPW}\&\ref{sec:IRW}, we can learn the optimal ranker given the correct propensity model and learn the optimal propensity model given the correct ranker.
Now we show that both of them can be achieved with the joint learning process of DLA.

For simplicity, we first consider the cases where $\theta$ is fixed for $S$.
%analyze the convergence of DLA with $\theta$ fixed for $S$.
Let $f_x$ and $g_x$ be the value of $P_S(r_q^x=1|\pi_q)$ and $P_E(o_q^x=1|\pi_q)$ in Equation~(\ref{equ:softmax}).
When $\theta$ is fixed, $\phi$ is the only learnable parameter in DLA.
As we only consider position bias in this work, $g_x = g_i$ ($i$ is the position of $x$ in $\pi_q$) and $\{g_i\}$ are independent with each other.
Suppose that $g_i$ is the softmax of $\phi_i$ over $\phi$ where $\phi_i$ is the $i$th column of $\phi$, then DLA will converge when: %$\frac{\partial \hat{\mathcal{L}}(E)}{\partial \phi_i} = 0$: %for $i \in [1, |\pi_q|]$:
\begin{equation}
\begin{split}
\frac{\partial \hat{\mathcal{L}}(E)}{\partial \phi_i} &= - \frac{1}{|Q|}\sum_{q\in Q}\frac{\partial \hat{l}_{IPW}(E,q)}{\partial \phi_i} \\
&= \frac{1}{|Q|}\sum_{q\in Q} \sum_{\substack{j=1}}^{|\pi_q|}\frac{c_q^jf_1}{f_j}g_i - \frac{c_q^if_1}{f_i} \\
&=g_i\sum_{\substack{j=1}}^{|\pi_q|}\frac{1}{|Q|}\sum_{q\in Q}\frac{c_q^jf_1}{f_j} - \frac{1}{|Q|}\sum_{q\in Q}\frac{c_q^if_1}{f_i} \\
&=g_i\sum_{\substack{j=1}}^{|\pi_q|}\mathbb{E}[\frac{c_q^jf_1}{f_j}] - \mathbb{E}[\frac{c_q^if_1}{f_i}] = 0
\end{split}
\end{equation}
%$\frac{\partial g_i}{\partial \phi_j} = - g_i, j\neq i$
We use the fact that $\frac{\partial g_i}{\partial \phi_j} = \bm{1}_{j=i} - g_j$ in step 2 and finally get that $g_i = \mathbb{E}[\frac{c_q^if_1}{f_i}] / \sum_{\substack{j=1}}^{|\pi_q|}\mathbb{E}[\frac{c_q^jf_1}{f_j}]$.
It worth noticing that $\hat{\mathcal{L}}(E)$ will always converge to its global minimum because it is concave:
$$
\frac{\partial^2 \hat{\mathcal{L}}(E)}{\partial \phi_i^2} = (1-g_i)\sum_{\substack{j=1}}^{|\pi_q|}\mathbb{E}[\frac{c_q^jf_1}{f_j}] \geq 0
$$
Therefore, when DLA converges, the inverse propensity weights produced by $E$ on position $i$ is 
\begin{equation}
\frac{g_1}{g_i} = \frac{\mathbb{E}[\frac{c_q^1f_1}{f_1}]}{\mathbb{E}[\frac{c_q^if_1}{f_i}]} = \frac{\mathbb{E}[c_q^1]}{\mathbb{E}[\frac{c_q^if_1}{f_i}]} = \frac{\mathbb{E}[o_q^1 \cdot r_q^1]}{\mathbb{E}[\frac{o_q^i \cdot r_q^i\cdot f_1}{f_i}]} = \frac{\mathbb{E}[\frac{r_q^1}{r_q^i}]}{\mathbb{E}[\frac{f_1}{f_i}]} \cdot \frac{\mathbb{E}[o_q^1] }{\mathbb{E}[o_q^i]}
\label{equ:convergence}
\end{equation}
%\frac{\mathbb{E}[o_q^1] \mathbb{E}[r_q^1]}{\mathbb{E}[o_q^i]\mathbb{E}[\frac{r_q^i\cdot f_1}{f_i \cdot r_q^1}]}
We use the fact that $c_q^i=o_q^i\cdot r_q^i$ and $\{o_q^i\}$, $\{r_q^i\}$, $\{f_i\}$ are independent given $\theta$.

In Equation~(\ref{equ:convergence}), $\mathbb{E}[r_q^1/r_q^i]$, $\mathbb{E}[f_1/f_i]$ are the real and estimated inverse relevance weights for $\hat{l}_{IRW}(E,q)$, and $\mathbb{E}[o_q^1]/\mathbb{E}[o_q^i] = P(o^1=1)/P(o^i=1)$ is the true inverse propensity weight we want to estimate for $\hat{l}_{IPW}(S,q)$.
%The goal of unbiased learning to rank is to learn an unbiased ranker $S$ such that $f_x = \mathbb{E}[r_q^x]$, so 
This indicates that \textit{the better the $S$ is as an unbiased ranker, the better the $E$ is as an unbiased propensity estimator}.
Similarly, we can prove its inverse proposition by fixing $\phi$ and deriving the derivative of $\theta$ with respect to $\hat{\mathcal{L}}(S)$.
As shown by McLachlan and Krishnan~\cite{mclachlan2007algorithm}, jointly optimizing two functions that control each other can converge to their global optima when both of them are concave. 
Because $\hat{\mathcal{L}}(E)$ is concave with respect to $\phi$, DLA will converge to the best unbiased ranker $S$ and propensity estimator $E$ when $\hat{\mathcal{L}}(S)$ is also concave with respect to $\theta$.
%the estimation performance of $E$ is controlled by the ranking performance of $S$.

% for simplicity, fix one and consider the convergen
% convergence -> gradient = 0
% we have xxxx = xxxx

% Given xxx, the estimated ipw is 
% because xx= xx*xx where xx and xx are independent, then we have xxx

% from xxx we can see that xxx is depend on xxx
% when xxx is better, xxx is better
% similarly, xxx depends on xxxx, when xxx is better xxx is better
% as long as xxx has the power to model xxx, DLA will converge to xxx

\subsection{Model Implementation} \label{sec:model_implementation}

Theoretically, any machine learning model that works with stochastic gradient decent (SGD) can be used to implement the ranker and propensity model in DLA.
% Implement of ranker
In this paper, we implement the ranker $S$ in DLA with deep neural networks (DNN).
Given a document's feature vector $\bm{x}$, we have
\begin{equation}
\begin{split}
\bm{h_{0}} &= \bm{x} \\
\bm{h_{k}} &= elu(\bm{W}_{k-1} \cdot \bm{h}_{k-1} + \bm{b}_{k-1}), k = 1,2,3,4
\end{split}
\label{equ:dnn}
\end{equation} 
where $\theta = \{\bm{W}_{k},\bm{b}_{k} | k = 0,1,2,3\}$ are the parameters learned from training data and $elu(x)$ is a non-linear activation function that equals to $x$ when $x \geq 0$ and $e^x-1$ otherwise.
The output of the network $h_4$ is a scalar, which will be used as the ranking score $f_q^x(\theta)$ for the document.

% implement of estimator
As we only consider position bias in this work, the most straightforward method to implement the propensity estimator $E$ is to represent the propensity score for each position with a separate variable.
%using separate variables $\{\phi_i\}$ to represent the propensity scores of each position $i$.
We tried other methods like converting positions into one-hot input vectors and using a DNN or recurrent neural network to predict the propensity scores.
However, we observe no benefit from applying these complicated models for $E$.
Thus, we only report the results of DLA that directly represents the propensity score of position $i$ with $\phi_i$.

%!TEX root=SIGIR18-NoisyLTR.tex
\section{Experimental Setup}\label{sec:setup}

To analyze the effectiveness of DLA, we conducted two types of experiments.
The first one is a simulation experiment based on a public learning-to-rank dataset.
The second one is a real-world experiment based on the actual ranked lists and user clicks collected from a commercial Web search engine.  

\subsection{Simulation Experiment Setup}\label{sec:simulation_setup}
%goal
%dataset
To fully explore the spectrum of click biases and the corresponding performance of DLA under different situations, we conducted experiments on the Yahoo! LETOR set 1\footnote{\url{http://webscope.sandbox.yahoo.com}} with derived click data.
Yahoo! LETOR data is one of the largest public learning-to-rank dataset from commercial English search engines.
In total, it contains 29,921 queries with 710k documents.
Each query-document pair has a 5-level relevance judgment and 700 features selected by a separate feature selection step in which the most predictive production features are kept~\cite{chapelle2011yahoo}.
We follow the same data split of training, valiation and testing in the Yahoo! LETOR set 1. 
Due to privacy concerns, no user information and click data was released with this dataset.
Thus, we need to sample synthetic click data for the training of unbiased learning-to-rank models.   

%data generation
\textbf{Click simulation}.
Similar to the setting used by Joachims et al.~\cite{joachims2017unbiased}, we generate click data on Yahoo! LETOR dataset with a two-step process.
First, we trained a Ranking SVM model~\cite{joachims2006training} using 1\% of the training data with real relevance judgments to generate the initial ranked list $\pi_q$ for each query $q$. 
We refer to this model as the \textit{Initial Ranker}.
Second, we sampled clicks on documents by simulating the browsing process of search users.
We assume that a user will click a search result ($c_q^x=1$) if and only if the document is observed ($o_q^x=1$) and perceived as relevant ($r_q^x=1$). 
To sample $o_q^x$, we adopted the presentation bias $\bm{\rho}$ estimated by Joachims et al.~\cite{joachims2005accurately} through eye-tracking experiments:
$$
P(o_q^x=1|\pi_q) = P(o_i=1) =  \rho_i^\eta
$$
where $\eta\in[0,+\infty]$ is a hyper-parameter that controls the severity of presentation biases.
We set $\eta=1.0$ if not discussed explicitly.
Following the methodology proposed by Chapelle et al.~\cite{chapelle2009expected}, we sampled $r_q^x$ with:
$$
P(r_q^x=1|\pi_q) = \epsilon + (1-\epsilon)\frac{2^y-1}{2^{y_{max}}-1}
$$
where $y \in [0,4]$ is the 5-level relevance label for document $x$ and $y_{max}$ is the maximum value of $y$ (which is 4 in our case).
We use a parameter $\epsilon$ to model click noise so that irrelevant documents ($y=0$) have non-zero probability to be perceived as relevant and clicked.
%We use a parameter $\epsilon$ to model the noise of users so that irrelevant documents ($y=0$) have non-zero probability to be perceived as relevant and thus to be clicked.
Joachims et al.~\cite{joachims2017unbiased} have proved that click noise does not affect the effectiveness of unbiased learning-to-rank with IPW framework as long as $P(r_q^x=1|\pi_q)$ is higher on relevant documents than irrelevant documents.
For simplicity, we fixed the value of $\epsilon$ as 0.1. 

%baseline
\textbf{Baselines}.
We included two groups of baselines in the simulation experiments.
The first group is the ranking models trained with click data directly without any bias correction, which is referred to as \textit{NoCorrect}.
%The goal is to understand the effect of different bias correction algorithms for unbiased learning to rank.
The second group is the existing unbiased learning-to-rank algorithms based on randomization experiments~\cite{joachims2017unbiased}, which is referred to as \textit{RandList}.
We randomly shuffle the results in the initial lists provided by the initial ranker and sampled 2 million click sessions to estimate the examination propensity on each position.
%For result randomizations on the initial lists provided by the initial ranker and sampled clicks to estimate the propensity model.
%We randomly collected 2 million click sessions to insure the quality of estimation.
%We refer to the strategy of not using bias correction as \textit{NoCorrect} and the strategy of using randomization-based bias correction as \textit{RandList}.
%cWe tested two types of rankers: IPW-based Ranking SVM~\cite{joachims2017unbiased} 
For ranking algorithms, we tested the Ranking SVM (which is used by Joachims et al.~\cite{joachims2017unbiased} in their initial study of unbiased learning to rank) and the deep neural network (DNN) described in Section~\ref{sec:model_implementation}.
In total, we have four baselines in our simulation experiments:
the Ranking SVM with NoCorrect/RandList and the DNN with NoCorrect/RandList.

%For the DNN, we implemented the ranking loss function as shown in Equation~\ref{equ:softmax_loss}.

% model training
\textbf{Model training}.
We trained all models with the training set of Yahoo! LETOR dataset based on synthetic clicks. 
Click sessions for training queries are sampled on the fly to avoid unnecessary biases introduced by off-line generations.
We used the Ranking SVM\footnote{\url{https://www.cs.cornell.edu/people/tj/svm_light/svm_proprank.html}} from Joachims et al.~\cite{joachims2017unbiased} and implemented the DNN model with Tensorflow\footnote{\url{https://www.tensorflow.org/}}.
We tuned the parameter $c$ from 20 to 200 for the Ranking SVM and tuned the number of hidden units from 128 to 512 for the DNN.
We trained the DNN with stochastic gradient descent and tuned the learning rate $\alpha$ from 0.005 to 0.05.
We set batch size as 256 and stopped training after 10k steps, which means that each model observed approximately 2.5 million click sessions.  
%The learning rate $\alpha$ are tuned from 0.005 to 0.05 and we conducted stochastic gradient descent with batch size 256. 
%All training process were stopped after 10,000 batches, which means that each model were trained with approximately 2.5 million click sessions.
In this paper, we only report the best results for each baseline.
Our code and synthetic data can be found in the following link\footnote{\url{https://github.com/QingyaoAi/Dual-Learning-Algorithm-for-Unbiased-Learning-to-Rank}}.

%metirc, evaluation method
\textbf{Evaluation}.
The evaluation of retrieval performance for baselines and DLA are conducted on the test set of Yahoo! LETOR data with expert judged relevance labels.
The evaluation metrics we used include the mean average precision (MAP), the normalized Discounted Cumulative Gain (nDCG)~\cite{jarvelin2002cumulated} and the Expected Reciprocal Rank (ERR)~\cite{chapelle2009expected}.
For both nDCG and ERR, we reported the results at rank 1, 3, 5 and 10 to show the performance of models on different positions.
Statistic differences are computed based on the Fisher randomization test~\cite{smucker2007comparison} with $p \leq 0.05$. 
We will discuss the results of the simulation experiments in Section~\ref{sec:simulation_results}.

% generate clicks on the fly to mimic online learning, avoid training bias

\subsection{Real-world Experiment Setup}

\begin{table}
	\caption{A summary of the ranking features extracted for our real-world experiments.}
	\small
	\def\arraystretch{1.15}%  1 is the default, change whatever you need
	\begin{tabular}
		{| l | p{0.355\textwidth}|} \hline
		TF & The average term frequency of query terms in url, title, content and the whole document.   \\\hline
		IDF & The average inverse document frequency of query terms in url, title, content and the whole document.   \\\hline
		TF-IDF & The average value of $tf\cdot idf$ of query terms in url, title, content and the whole document.   \\\hline
		BM25 & The scores of BM25~\cite{robertson1994some} on url, title, content and the whole document.   \\\hline
		LMABS & The scores of Language Model (LM)~\cite{ponte1998language} with absolute discounting~\cite{zhai2017study} on url, title, content and the whole document.   \\\hline
		LMDIR & The scores of LM with Dirichlet smoothing~\cite{zhai2017study} on url, title, content and the whole document.   \\\hline
		LMJM & The scores of LM with Jelinek-Mercer~\cite{zhai2017study} on url, title, content and the whole document.   \\\hline
		Length & The length of url, title, content and the whole document.   \\\hline
		Slash & The number of slash in url.   \\\hline
	\end{tabular}\label{tab:features}
	%\vspace{-10pt}
\end{table}
%goal
%dataset
In order to show the effectiveness of DLA for unbiased learning to rank in practice, we collected click data from a commercial Web search engine.
We randomly sampled 3,449 queries written by real search engine users and collected the top 10 results from a two-week search log.
We downloaded the raw HTML documents based on urls and removed ranked lists which contain documents that cannot be reached by our crawler.
%We also removed the documents with less than 10 clicks 
After cleaning, we have 333,813 documents, 71,106 ranked lists and 3,268,177 anonymized click sessions in total.

%data collection
\textbf{Feature extraction}.
For the training of learning-to-rank algorithms, we manually extracted features based on the text of queries and documents.
Following a similar methodology used by Microsoft Letor data~\cite{qin2010letor}, we designed features based on url, title, content and the whole text of the documents.
In total, we have 33 features for each query-document pair.
The ranking features used in our experiments are summarized in Table~\ref{tab:features}.

%\footnote{\url{https://www.microsoft.com/en-us/research/project/letor-learning-rank-information-retrieval/}}

%baseline
\textbf{Baselines}.
Due to the limits of our access to the commercial system, we cannot conduct result randomization experiments on real users.
Therefore, we focus on the comparison of DLA with other bias correction methods built on user clicks.
More specifically, we compared our approach with the model trained with relevance signals estimated by click models~\cite{dupret2008user,chapelle2009dynamic}. 
%Therefore, we focus on the comparison of DLA with other bias correction methods built on user clicks.
Click models are designed to extract the true relevance feedback from click data through making hypotheses on user browsing behaviors.
In our experiments, we implemented two click models: the user browsing model (UBM)~\cite{dupret2008user} and the dynamic bayesian network model (DBN)~\cite{chapelle2009dynamic}.
UBM assumes that the examination of a document depends on its position and its distance to the last click.
DBN assumes that users will keep reading documents sequentially and click them if they look attractive. 
If not satisfied, users will have a constant probability to return to the search result page and continue reading.
Both UBM and DBN use the Expectation-Maximization algorithm~\cite{mclachlan2007algorithm} to estimate their parameters based on click logs.
To insure the quality of relevance estimation with click models, we removed ranked lists with less than 10 click sessions in the cleaning process described previously.
%The final baseline models are trained with the relevance signals extracted by UBM and DBN using the same DNN model as described in Section~~\ref{sec:model_implementation}.
The final baselines are the DNN models (in Section~~\ref{sec:model_implementation}) trained with the relevance signals extracted by UBM and DBN.
Other training settings are the same as those used in the simulation experiments.

%\textbf{Model training}.

\textbf{Evaluation}.
The evaluation of unbiased learning-to-rank algorithms requires human judgments of query-document relevance.
In our experiments, we constructed a separate test dataset with 100 queries and recruited professional assessors to judge the relevance of top 100 documents retrieved by BM25~\cite{robertson1994some} in five level (\textit{irrelevant}, \textit{fair}, \textit{good}, \textit{excellent} and \textit{perfect}).
We trained our models and baselines on the training set with clicks and evaluated their performance on the test set with human annotations.
Similar to the simulation experiments, we reported the value of MAP, nDCG and ERR for all models in Section~\ref{sec:real-world_results}.
Please refer to \cite{luo2018sogou} for the data used in this paper.
%We plan to release our code as well as the training and testing data after the paper is published.

%!TEX root=SIGIR18-NoisyLTR.tex
\begin{table*}[ht]
	\centering
	\small
	\caption{Comparison of different unbiased learning-to-rank models on Yahoo! LETOR set 1. Significant improvements or degradations with respect to the DNN with DLA are indicated with $+/-$. %in the Fisher randomization test~\cite{smucker2007comparison} with $p \leq 0.01$. The best performance is highlighted in boldface.
	}
	%\scalebox{0.9}{
	\begin{tabular}{ c | c || c | c | c | c | c | c | c | c | c    } %p{5mm}
		\hline
		%\multicolumn{1}{c||}{ } & \multicolumn{3}{c||}{Dataset 1} & \multicolumn{3}{c}{Dataset 2}\\ \hlineB{3}
		\multicolumn{11}{c}{Yahoo! LETOR set 1}\\ \hline 
		Ranking Model & Correction Method& MAP & nDCG@1 & ERR@1 & nDCG@3 & ERR@3 & nDCG@5 & ERR@5 & nDCG@10 & ERR@10  \\\hline \hline
		\multicolumn{2}{c||}{DNN with DLA} & 0.816 & 0.658 & 0.338 & 0.662 & 0.412 & 0.683 & 0.433 & 0.729 & 0.447 \\ \hline \hline
		\multirow{ 2}{*}{Ranking SVM} & NoCorrect  & 0.814$^-$ & 0.628$^-$ & 0.316$^-$ & 0.638$^-$ & 0.395$^-$ & 0.661$^-$ & 0.416$^-$ & 0.711$^-$ & 0.432$^-$ \\ \cline{2-11}
		& RandList & 0.812$^-$ & 0.642$^-$ & 0.330$^-$ & 0.653$^-$ & 0.407$^-$ & 0.675$^-$ & 0.428$^-$ & 0.721$^-$ & 0.442$^-$ \\ \hline \hline
		\multirow{ 2}{*}{DNN} &  NoCorrect & 0.807$^-$ & 0.622$^-$ & 0.317$^-$ & 0.631$^-$ & 0.394$^-$ & 0.653$^-$ & 0.416$^-$ & 0.704$^-$ & 0.431$^-$\\ \cline{2-11}
		& RandList & 0.814$^-$ & 0.658 & 0.338 & 0.659$^-$ & 0.412 & 0.679$^-$ & 0.433 & 0.725$^-$ & 0.447 \\ \hline \hline
		\multicolumn{2}{c||}{Initial Ranker} & 0.804$^-$ & 0.559$^-$ & 0.271$^-$ & 0.586$^-$ & 0.357$^-$ & 0.617$^-$ & 0.381$^-$ & 0.675$^-$ & 0.397$^-$ \\ \hline
		\multicolumn{2}{c||}{Oracle DNN} & 0.830$^+$ &	0.667$^+$ & 0.339$^+$ & 0.675$^+$ & 0.414$^+$ & 0.695$^+$ & 0.435$^+$ & 0.740$^+$ & 0.449$^+$ \\ \hline
		
	\end{tabular}
	%}
	\vspace{-5pt}
	\label{tab:simulation}
\end{table*}

\section{Results and analysis}\label{sec:results}

In this section, we discuss the results of our simulation experiments and real-world experiments.
In particular, we focus on the following research questions:
\begin{itemize}
\item \textbf{RQ1}: Can DLA effectively estimate the true presentation bias and produce an unbiased ranker at the same time?
\item \textbf{RQ2}: Compared to the methodology that debiases click data and trains learning-to-rank models separately, are there any benefits from the joint learning of rankers and propensity models empirically?
\end{itemize}

\subsection{Comparison with Result Randomization}\label{sec:simulation_results}

To answer RQ1, we compare DLA with the unbiased learning-to-rank algorithms built on result randomization in the simulation experiments.
Specifically, we consider two scenarios.
In the first scenario, we generated the synthetic clicks with a single bias model $\bm{\rho}$ in result randomization and model training.
In the second scenario, we fixed $\bm{\rho}$ in result randomization but disturb its severity parameter $\eta$ in model training.
We refer to the first scenario as the \textit{Oracle Mode} and the second scenario as the \textit{Realistic Mode}.

\textbf{Oracle Mode}.
The motivation of Oracle Mode is to test unbiased learning-to-rank algorithms in cases where click bias does not change over time. 
The performance of ranking models trained with different bias correction methods in this scenario is summarized in Table~\ref{tab:simulation}.
%\textit{NoCorrect} and \textit{RandList} denote training with raw clicks and bias correction based on randomization experiments respectively.  
For better illustration, we also include the results of the initial ranked lists (Initial Ranker) and the DNN model trained with human annotations (Oracle DNN).

% clicks are good without correction.
As shown in Table~\ref{tab:simulation}, feedback information from click data indeed helped improve the performance of ranking models.
Even with no bias correction, the performance of Ranking SVM and DNN trained with click data are better than the Initial Ranker.
% DNN perform better that SVM ranking
Comparing the two ranking algorithms, the DNN with softmax cross entropy consistently outperformed Ranking SVM.
% Randlist performs bettern than no correction
After incorporating bias corrections with the propensity model estimated by result randomization (RandList), we observed approximately 3\% improvements with respect to ERR@10 on Ranking SVM and DNN over the models trained with raw clicks.
This demonstrated the effectiveness of unbiased learning to rank with inverse propensity weighting and RandList.
In Oracle Mode, we manually ensured that the presentation bias in the randomization experiments is the same as those in the training data.
As discussed in Section~\ref{sec:randomization}, result randomization is guaranteed to find the true click propensity in theory~\cite{joachims2017unbiased}.
Therefore, the ranking models trained with RandList can be treated as the optimal ranker we can get with unbiased learning-to-rank algorithms.

% DLA is similar with Rand list and approaching oracle
Comparing the DNN models trained with DLA and RandList, we find that DLA is as effective as (if not better than) RandList in terms of bias correction.
The DNN with DLA performed similarly with the DNN with RandList and was significantly better than other baselines.  
Its performance is close to the Oracle DNN, which is trained with human relevance judgments.
% DLA is at least no worse and doesn't require randominzation, preferable
Because the DNN with DLA does not use result randomization and performed as effectively as the model trained with RandList, it has advantages in real retrieval applications.
%By automatically estimating propensity from training data, DLA naturally adapts to the change of presentation bias.
%We will give more analysis on its adaptiveness as follows.

\textbf{Realistic Mode}. 
The assumption of Oracle Mode that click biases remain unchanged in randomization experiments and model training is not realistic.
Because we frequently introduce new features into search engine interfaces, user behaviors evolve rapidly and the propensity model estimated by result randomization can be out-of-date and inconsistent with the true click bias.
To model such scenarios, we fixed $\eta$ as 1 for click sampling in result randomization and used different $\eta$ to generate clicks for model training. 

\begin{figure}[t]
	%\vspace{-5pt}
	\centering
	\includegraphics[width=2.4in]{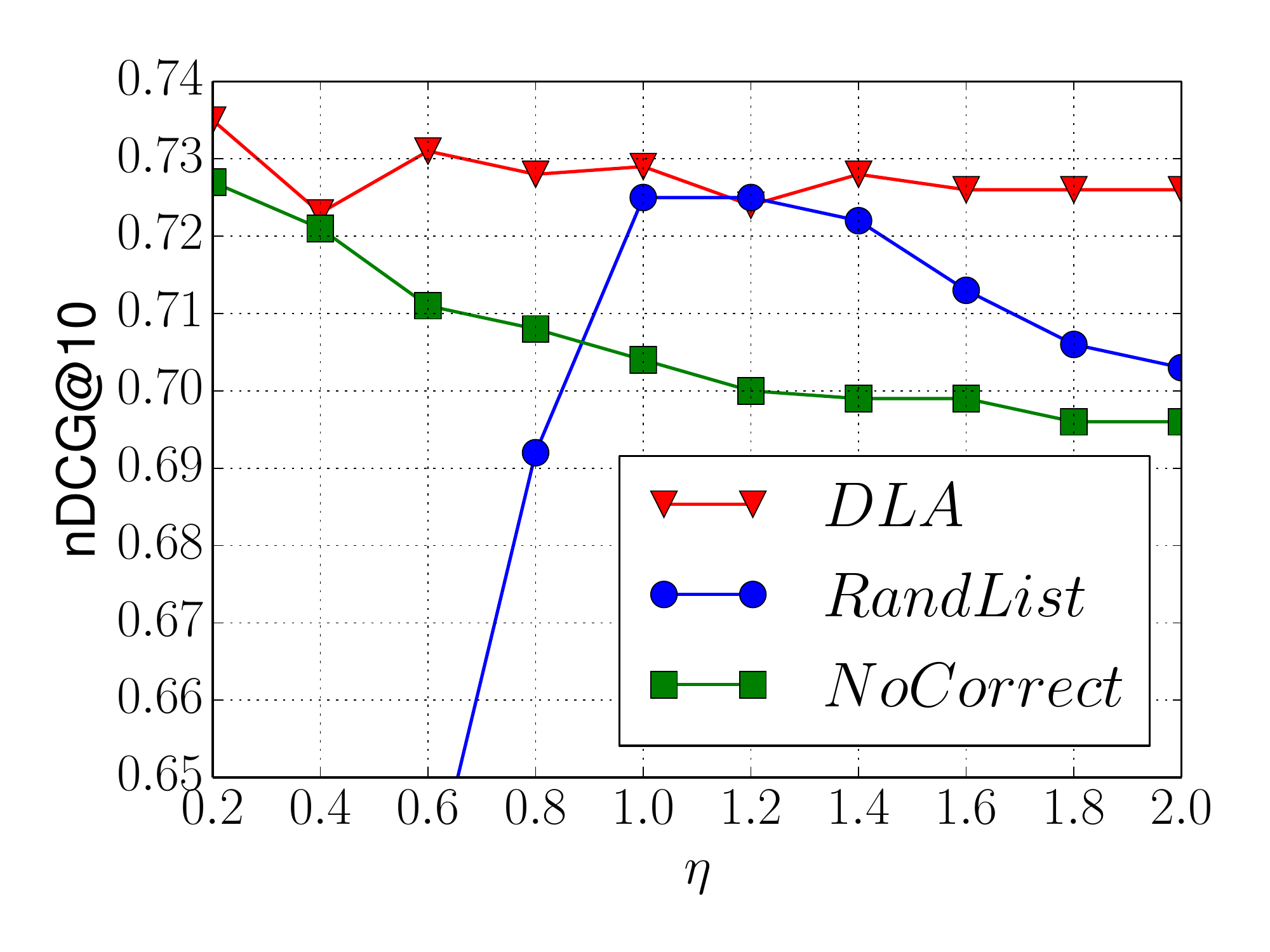}
	\setlength{\belowcaptionskip}{-10pt}
	\setlength{\abovecaptionskip}{-5pt}
	\caption{The performance of DNN trained with different bias corrections with respect to the value of $\eta$.}
	\label{fig:eta}
	%\vspace{-5pt}
\end{figure}

% plot
Figure~\ref{fig:eta} depicts the performance of the DNN models trained with NoCorrect, RandList and DLA with respect to different $\eta$. 
% explain eta, small, large
The presentation bias is severe when $\eta$ is large and vanishes when $\eta=0$.
% no correct 
As shown in Figure~\ref{fig:eta}, the performance of the DNN with NoCorrect is negative correlated with the value of $\eta$.
Without bias correction, the training of ranking models are exposed to click bias.
Thus, an increase of $\eta$ will hurt the performance of models trained with raw clicks.
Compared to NoCorrect, the effect of misspecified $\eta$ on the DNN with RandList is more complicated.
When $\eta$ is 1 (which is same with the click sampling process used in result randomization), the DNN with RandList outperformed the DNN with NoCorrect.
When $\eta>1$, the relative improvements of RandList over NoCorrect are positive but keep decreasing as $\eta$ increases.
Although the models with RandList underestimated the real presentation bias, they are still better than those with no bias correction.    
When $\eta < 1$, however, the DNN with RandList performed poorly and are worse than the DNN with NoCorrect.
When click biases in the training data are not as severe as they are in the randomization experiments, RandList overestimated the real biases and introduced extra noise into the training of ranking models.
In comparison, the performance of the DNN with DLA is robust to changes of $\eta$ and significantly outperformed NoCorrect and RandList in most cases.
Because it automatically and directly estimates propensity model from training data, DLA is adaptive to changes in click biases and more robust in practice.

\begin{figure}[t]
	%\vspace{-5pt}
	\centering
	\includegraphics[width=2.4in]{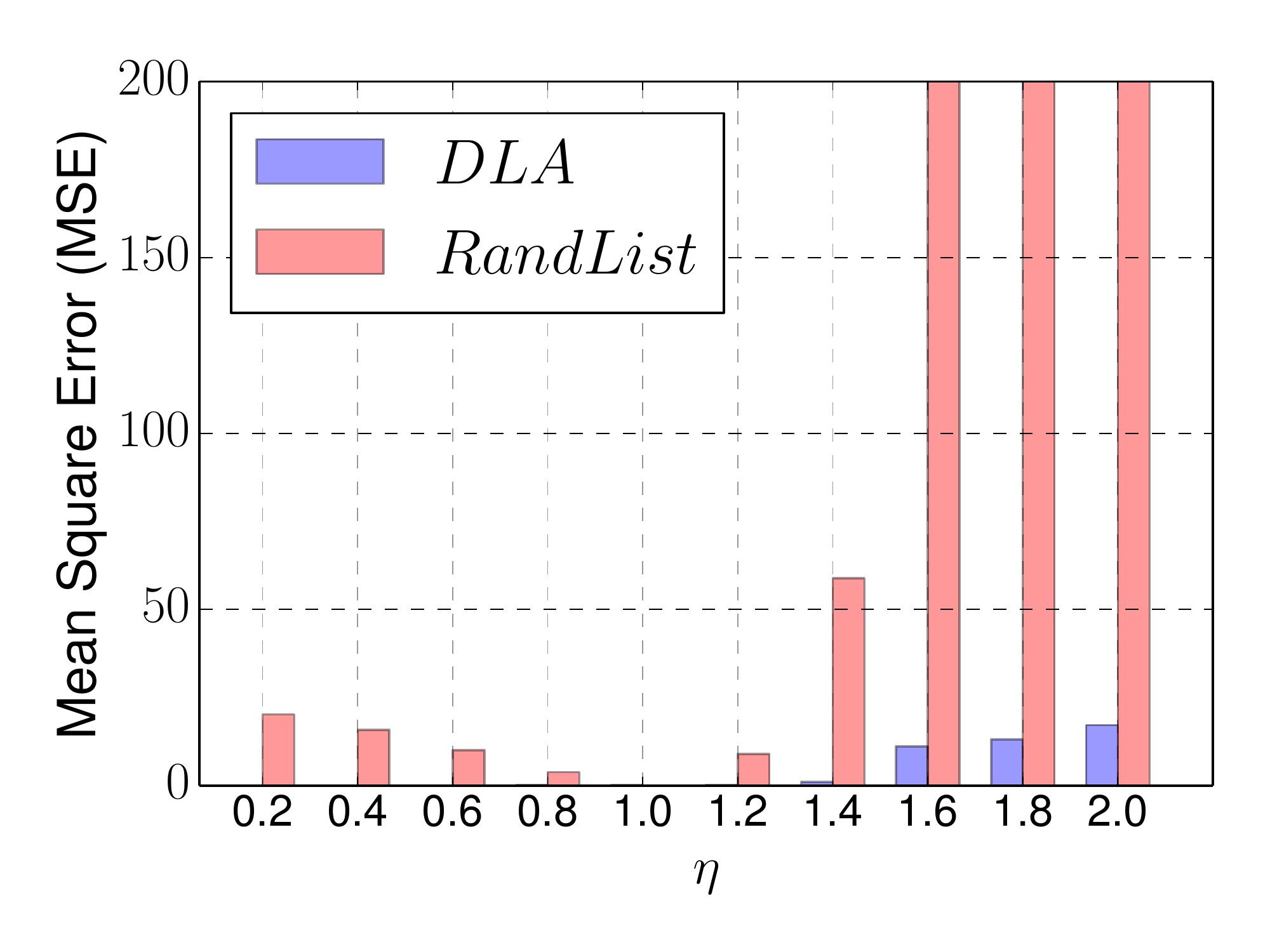}
	\setlength{\belowcaptionskip}{-5pt}
	\setlength{\abovecaptionskip}{-5pt}
	\caption{The MSE between the true click propensity and those estimated by DLA and RandList with respect to $\eta$.}
	\label{fig:exam}
	%\vspace{-5pt}
\end{figure}

% connection
To explain why DLA produced a better unbiased ranker than RandList, we computed the mean square error (MSE) between the true inverse propensity weights ($\rho_0^\eta/\rho_i^\eta$) and the inverse propensity weights ($g_0/g_i$) estimated by DLA or RandList:
$$
MSE = \frac{1}{|\pi_q|} \sum_{i=0}^{|\pi_q|-1} \big(\frac{g_0}{g_i} - \frac{\rho_0^\eta}{\rho_i^\eta}\big)^2
$$ 
As shown in Figure~\ref{fig:exam}, the MSE of RandList is small when $\eta=1$ but large otherwise. %, especially when $\eta$ is greater than 1.
In contrast, the MSE of DLA is small for all $\eta$.
This indicates that the propensity model estimated by DLA is always better for approaching the true presentation bias in spite of $\eta$.
As discussed in Section~\ref{sec:IPW}, the estimation of inverse propensity weights is the key for unbiased learning to rank~\cite{wang2016learning,joachims2017unbiased}.
Therefore, the models trained with DLA outperformed the models trained with RandList significantly and consistently.
% explain plot
% RandList not work
% our model works.

% Randlist not well, smaller eta overestimated, larger eta, under estimated
% The DLA consistently perform well.

% we may add online experiments if necessary

\subsection{Comparison with Click Models}\label{sec:real-world_results}

\begin{table*}[t]
	\centering
	\small
	\caption{Comparison of DNN trained with DLA and relevance signals extracted by click models. Significant improvements or degradations with respect to DLA are indicated with $+/-$. % in the Fisher randomization test~\cite{smucker2007comparison} with $p \leq 0.01$. The best performance is highlighted in boldface.
	}
	%\scalebox{0.9}{
	\begin{tabular}{  c || c | c | c | c | c | c | c | c | c    } %p{5mm}
		\hline
		%\multicolumn{1}{c||}{ } & \multicolumn{3}{c||}{Dataset 1} & \multicolumn{3}{c}{Dataset 2}\\ \hlineB{3}
		Correction Method & MAP & nDCG@1 & ERR@1 & nDCG@3 & ERR@3 & nDCG@5 & ERR@5 & nDCG@10 & ERR@10  \\\hline \hline
		DLA & 0.881 & 0.433 & 0.406 & 0.410 & 0.537 & 0.422 & 0.571 & 0.421 & 0.582 \\ \hline\hline
		DBN & 0.865$^-$ & 0.363 & 0.340 & 0.370 & 0.468$^-$ & 0.390 & 0.504$^-$ & 0.419 & 0.521$^-$ \\ \hline
		UBM & 0.849$^-$ & 0.359$^-$ & 0.336$^-$ & 0.343$^-$ & 0.464$^-$ & 0.352$^-$ & 0.502$^-$ & 0.365$^-$ & 0.519$^-$ \\ \hline \hline
		NoCorrect & 0.810$^-$ & 0.357$^-$ & 0.334$^-$ & 0.348$^-$ & 0.459$^-$ & 0.349$^-$ & 0.484$^-$ & 0.358$^-$ & 0.500$^-$ \\ \hline
	\end{tabular}
	\vspace{-5pt}
	%}
	\label{tab:real-world}
\end{table*}

To answer RQ2 and to demonstrate the effectiveness of DLA on real click data, we compare the performance of the DNN models trained with DLA and click models in the real-world experiment.
A summary of the results is shown in Table~\ref{tab:real-world}.
%To better illustrate the effect of bias correction, we also show the results of the DNN trained with raw user clicks as baselines.

As we can see from Table~\ref{tab:real-world}, the ranking models trained with click model signals (DBN and UBM) consistently outperformed the model trained with raw clicks (NoCorrect).
This empirically demonstrates that click models can extract better relevance signals from user clicks and are beneficial for the learning of unbiased rankers. 
Among the two click models tested in our experiments, the ranker trained with DBN performed better than the ranker trained with UBM. 
If we compare DLA with click models, we can see that the model trained with DLA achieved significantly improvements over the models trained with DBN and UBM.
It obtained more than 10\% improvements over DBN and UBM in terms of ERR@10.
  
The fact that the DNN model trained with DLA produced better results than those trained with click models indicates that the joint learning of bias correction and ranking models is promising for unbiased learning to rank.
First, the joint learning paradigm simplifies the design of search engines as it directly utilizes raw click data without requiring separate experiments to debias user feedback.
This is important because such experiments are either expensive or have special requirements for data (e.g. click models require each query-document pair to appear multiple times).
Second, as discussed in Section~\ref{sec:convergence}, the learning of propensity models and ranking models can help each other.
A propensity model jointly learned with rankers observes more document information in the ranked lists, so it has a better chance to estimate the true click bias. 

Last but not least, joint learning enables DLA to conduct end-to-end optimization for unbiased learning to rank.
%DLA is an end-to-end algorithm that directly optimizes the performance of unbiased learning to rank.
In fact, an important drawback of existing learning paradigms for unbiased learning to rank is that they optimize different objectives in different learning steps.
For example, most click models are designed to optimize the likelihood of observed clicks.
Although Malkevich et al.~\cite{Malkevich:2017:EAC:3121050.3121096} have shown that UBM is better than DBN in terms of click simulation, the ranker trained with DBN performed better than the ranker trained with UBM in our experiments.
This suggests that optimizing click likelihood doesn't necessarily produce the best bias correction model for unbiased learning to rank.
Even though we implemented the propensity model with a simple examination hypothesis compared to UBM and DBN, the ranker trained with DLA still significantly outperformed the rankers trained with the click models.
This demonstrates the benefits of end-to-end training with the joint learning paradigm.

%On the other hand

%The joint learning in DLA is an end-to-end process so that we can directly 

% click modeling may not be relevance modeling

%\input{discussion}

%!TEX root=SIGIR18-NoisyLTR.tex
\section{Conclusion and Future Work}\label{sec:conclusion}

In this work, we propose a Dual Learning Algorithm for automatic unbiased learning to rank.
DLA jointly learns unbiased propensity models and ranking models from user clicks without any offline parameter estimation or online result randomization.
Our analysis and experiments show that DLA is an theoretically principled and empirically effective framework for unbiased learning to rank.
This indicates that jointly learning propensity models and ranking models could be a fruitful direction for learning to rank with biased training signals.

Our work represents an initial attempt for automatic unbiased learning to rank and there are still many problems to study in this field.
For example, as shown in Section~\ref{sec:convergence}, the performance of propensity estimation depends on the quality of the ranking model. 
It seems that DLA does not work well when the best ranker we can get has poor performance.
We leave the investigation of these problems for future studies.
%Whether DLA will work when the best ranker we can get is far from perfect is still an open question.

% future work
% DLA not work when the best ranker is poor

\iftrue
\section{Acknowledgments}
This work was supported in part by the Center for Intelligent Information Retrieval. Any opinions, findings and conclusions or recommendations expressed in this material are those of the authors and do not necessarily reflect those of the sponsor.
\fi

\balance
\bibliographystyle{ACM-Reference-Format}
\bibliography{sigproc} 

\end{document}